\documentclass[PAPER, cernpreprint, atlasdraft=false, USenglish, texmf, orcidlogo]{atlasdoc}
 
\usepackage{atlaspackage}
 
\usepackage{atlasbiblatex}
 
\usepackage{atlasphysics}

\graphicspath{{logos/}{figures/}}
 
\usepackage{ANA-HDBS-2021-18-PAPER-defs}
\usepackage{multirow}
\usepackage{tabularx}
\usepackage{comment}
\usepackage{bm}
\usepackage[compat=1.1.0]{tikz-feynman}
\tikzfeynmanset{warn luatex=false}
\definecolor{hh_pink}{HTML}{F2385A}
\definecolor{hh_blue}{HTML}{343844}
\definecolor{hh_med_turq}{HTML}{36B1BF}
\definecolor{hh_lgt_turq}{HTML}{4AD9D9}
\definecolor{hh_whte}{HTML}{E9F1DF}
\definecolor{hh_yllw}{HTML}{CA9E2B}
\definecolor{hh_gren}{HTML}{00CC00}

\AtlasTitle{Combination of Searches for Higgs Boson Pair Production in $pp$ Collisions at $\sqrt{s}=13$~TeV with the ATLAS Detector}

\AtlasAbstract{
This Letter presents results from a combination of searches for Higgs boson pair production using 126--140 fb$^{-1}$ of proton-proton collision data at $\sqrt{s}=13$~TeV recorded with the ATLAS detector.
At 95\% confidence level (CL), the upper limit on the production rate is 2.9 times the standard model (SM) prediction, with an expected limit of 2.4 assuming no Higgs boson pair production. Constraints on the Higgs boson self-coupling modifier $\kappa_{\lambda}=\lambda_{HHH}/\lambda_{HHH}^\mathrm{SM}$, and the quartic $HHVV$ coupling modifier $\kappa_{2V}=g_{HHVV}/g_{HHVV}^\mathrm{SM}$, are derived individually, fixing the other parameter to its SM value.
The observed 95\% CL intervals are $-1.2 < \kappa_{\lambda} < 7.2$ and $0.6 < \kappa_{2V} < 1.5$, respectively, while the expected intervals are $-1.6 < \kappa_{\lambda} < 7.2$ and $0.4 < \kappa_{2V} < 1.6$ in the SM case. Constraints obtained for several interaction parameters within Higgs effective field theory are the strongest to date, offering insights into potential deviations from SM predictions.
}
 
\author{ATLAS Collaboration}
 
\AtlasRefCode{HDBS-2021-18}
 
\PreprintIdNumber{CERN-EP-2024-160}

\AtlasJournalRef{Phys. Rev. Lett. 133 (2024) 101801}
\AtlasDOI{10.1103/PhysRevLett.133.101801}

\hypersetup{pdftitle={ATLAS document},pdfauthor={The ATLAS Collaboration}}
 
\begin{document}
 
\maketitle

\label{sec:intro}

 
Since the discovery of the Higgs boson ($H$) by the ATLAS and CMS Collaborations at the Large Hadron Collider (LHC) in 2012~\cite{HIGG-2012-27,CMS-HIG-12-028}, understanding its intrinsic properties and interactions has been a priority.
The Higgs self-coupling is directly related to the shape of the Higgs scalar field potential, which is important for understanding the mechanism of electroweak symmetry breaking and serves as an essential test of the electroweak theory. After the symmetry breaking, the Higgs potential predicted in the standard model (SM) can be expanded in the Higgs boson field $H$ near its minimum: $V(H) = \frac{1}{2}m_H^2 H^2 + \lambda_{HHH}vH^3 + \mathcal{O}(H^4)$, where $m_{H}$ is the Higgs boson mass and $v\approx 246~\GeV$ is the field's vacuum expectation value~\cite{Zyla:2020zbs}.
The Higgs boson's trilinear self-coupling $\lambda_{HHH}^\mathrm{SM}$ is equal to $m_H^2/2v^2$, its coupling $g_{HVV}^\mathrm{SM}$ to vector bosons $(V=W,Z)$ is equal to $2m_{V}^{2}/v$, and its coupling $g_{Hf\bar{f}}^\mathrm{SM}$ to fermions is equal to $m_{f}/v$. The quartic coupling between two Higgs bosons and two vector bosons, $g_{HHVV}^\mathrm{SM}$, is equal to $g_{HVV}^\mathrm{SM}/v$~\cite{Zyla:2020zbs}.
The production of Higgs boson pairs ($HH$) via gluon-gluon fusion (ggF) and vector-boson fusion (VBF) provides a direct probe of $\lambda_{HHH}$ and $g_{HHVV}$, which affects the pair-production differential cross section at tree level.
Observed (expected) 95\% confidence level (CL) upper limits on the \HH production cross section have been set at 2.4 (2.9) and 3.4 (2.5) times the SM prediction by previous ATLAS~\cite{HDBS-2022-03} and CMS~\cite{CMS-HIG-22-001} search combinations, respectively.
 
\begin{figure}[tb]
\centering
\subfloat[]{\includegraphics[width=0.48\textwidth]{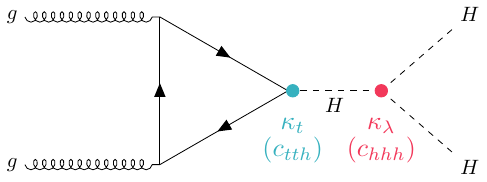}\label{fig:feyn_hha}}
\hspace{0.03\textwidth}
\subfloat[]{\includegraphics[width=0.48\textwidth]{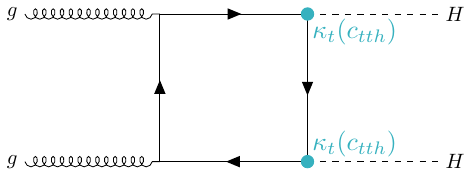}\label{fig:feyn_hhb}}\\
\subfloat[]{\includegraphics[width=0.32\textwidth]{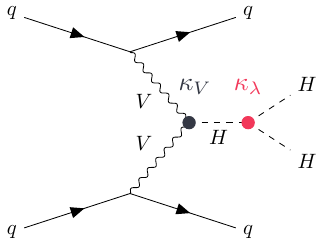}\label{fig:feyn_hhc}}
\hspace{0.013\textwidth}
\subfloat[]{\includegraphics[width=0.32\textwidth]{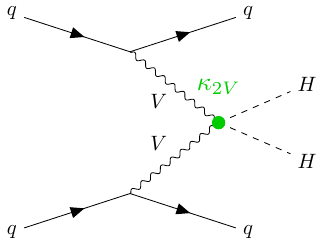}\label{fig:feyn_hhd}}
\hspace{0.013\textwidth}
\subfloat[]{\includegraphics[width=0.32\textwidth]{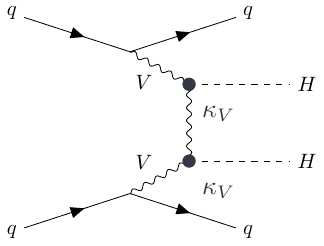}\label{fig:feyn_hhe}}\\
\subfloat[]{\includegraphics[width=0.32\textwidth]{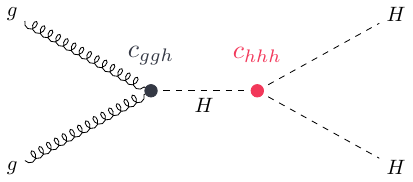}\label{fig:feyn_hhf}}
\hspace{0.033\textwidth}
\subfloat[]{\includegraphics[width=0.24\textwidth]{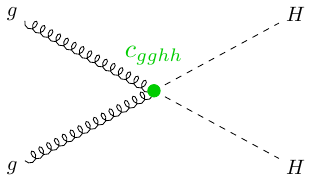}\label{fig:feyn_hhg}}
\hspace{0.033\textwidth}
\subfloat[]{\includegraphics[width=0.36\textwidth]{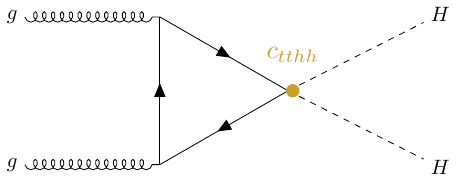}\label{fig:feyn_hhh}}\\
\caption{Leading-order Feynman diagrams showing the production of Higgs boson pairs via the ggF (a), (b), (f)--(h) and VBF (c)--(e) processes. Each diagram is sensitive to specific coupling factors, denoted by $\kappa_{i}$ in the $\kappa$ framework or $c_{i}$ in the HEFT. Diagrams (a)--(e) occur in the SM predictions, while diagrams (f)--(h) manifest only when deviations from the SM predictions are present in the coefficients \cggh, \cgghh, or \ctthh.}
\label{fig:feyn_hh}
\end{figure}

Deviations from the SM can be expressed in terms of coupling modifiers $\kappa_{i}$~\cite{Heinemeyer:1559921, deFlorian:2016spz} or Wilson coefficients $c_{i}$ in the Higgs effective field theory (HEFT)~\cite{ALONSO2013330, BUCHALLA2014552}, as illustrated in Fig.~\ref{fig:feyn_hh}.
In the $\kappa$ framework, the coupling modifiers \kl, \kt, \kv, and \kvv are each defined as the ratio of the Higgs boson coupling to its SM value, e.g., $\kl = \lambda_{HHH}/\lambda_{HHH}^\mathrm{SM}$.
In the HEFT framework, new physics in the electroweak sector is described through anomalous couplings of the Higgs boson. The organization of the HEFT Lagrangian is guided by chiral perturbation theory, with the low-energy dynamics of electroweak symmetry breaking described using a nonlinear realization of the gauge symmetry group $\mathrm{SU}(2)_{L}\times \mathrm{U}(1)_{Y}$. One advantage of the HEFT framework is that the anomalous single-Higgs-boson and \HH couplings are defined separately, allowing simplified \HH\ interpretations.
In the HEFT Lagrangian, ggF \HH production is described at leading order by five relevant operators, and their associated Wilson coefficients are \ctth, \cggh, \chhh, \cgghh, and \ctthh.
In this formalism, \ctth and \chhh are equivalent to \kt and \kl.
In this analysis, \kv, \kt (\ctth), and \cggh are set equal to the SM predictions, because those parameters are constrained by precise measurements of single-Higgs-boson production~\cite{HIGG-2023-11, CMS-HIG-22-001}.
 
In the SM, destructive interference between the ggF \HH production diagrams in Figs.~\ref{fig:feyn_hha} and~\ref{fig:feyn_hhb} makes softer Higgs bosons more sensitive for constraining \kl.
For $m_H=125$~\GeV and $\sqrt{s}=13$~\TeV proton-proton collisions, the predicted cross section is $\sigma_{\textrm{ggF}}^\textrm{SM}(HH) = 31.1^{\,+1.9}_{-7.1}\,(\textrm{scale} + m_{\textrm{top}}) \pm{0.9}\,(\textrm{PDF}+\alphas)$~fb~\cite{Dawson_1998,Borowka:2016ehy,Baglio_2019,de_Florian_2013,Shao_2013,deflorian2015higgs,Grazzini_2018,Baglio_2021} at next-to-next-to-leading order in \alphas and including and including an uncertainty related to the choice of the virtual top-quark mass scheme~\cite{Baglio_2021}.
The \enquote{$\textrm{PDF}+\alphas$} uncertainty accounts for uncertainties in the parton distribution functions and strong coupling constant, the \enquote{scale} uncertainty is due to the finite order of quantum chromodynamics (QCD) calculations, and the  \enquote{$m_{\textrm{top}}$} uncertainty is related to the top-quark mass scheme.
For SM VBF \HH production, divergences in the diagrams shown in Figs.~\ref{fig:feyn_hhd} and~\ref{fig:feyn_hhe} cancel out due to perturbative unitarity. If
\kvv deviates from the SM prediction, this cancellation no longer occurs, leading to a linear dependence of the cross section on the effective center-of-mass energy of the incoming vector bosons~\cite{vbf-pheno}. Consequently, the Higgs bosons are expected to be more energetic in non-SM scenarios.
The cross section for VBF \HH production is $\sigma_{\mathrm{VBF}}^\textrm{SM}(HH) = 1.73\pm0.04$~fb at next-to-next-to-next-to-leading order in QCD~\cite{Baglio_2013,Frederix_2014,Liu-Sheng:2014gxa,Dreyer:2018qbw,Dreyer:2018rfu}.

This Letter presents a combination of results from the \bbbb~\cite{HDBS-2019-29, HDBS-2022-02}, \bbtt~\cite{HDBS-2019-27}, \bbyy~\cite{HDBS-2021-10}, \ml~\cite{HDBS-2019-04}, and \bbll~\cite{HDBS-2019-02} decay channels, probing more than half of the \HH decays.
The first three analyses have been improved since the previous combination~\cite{HDBS-2022-03}, and the other two are newly included.
The \HHbbbb decay mode has the advantage of having the largest SM \HH\ decay branching fraction (33.9\%), but it also has the largest SM background, due to the abundance of QCD multijet events. Given its capability to probe relatively high-energy Higgs bosons, both the resolved~\cite{HDBS-2019-29} and boosted topologies~\cite{HDBS-2022-02} are now used to reconstruct the Higgs bosons.
The \HHbbtt decay mode has one of the larger branching fractions (7.3\%) among the investigated \HH decay channels and benefits from having only moderate background contamination. In the corresponding search~\cite{HDBS-2019-27}, one of the $\tau$ leptons is required to decay hadronically, ensuring orthogonality with the \bbll search.
Although the \HHbbyy decay mode has a small branching fraction (0.26\%), it has high trigger efficiency and a clean experimental signature.
The \bbtt~\cite{HDBS-2019-27} and \bbyy~\cite{HDBS-2021-10} analyses have been improved through optimized classification of selected events to enhance the sensitivity to the Higgs boson couplings. Furthermore, the \bbtt analysis now benefits from more accurate background modeling and larger samples of simulated events.
The \ml analysis is designed to select \HH events in \bbZZ, \VVVV ($V=W\text{ or }Z$), \VVtt, \tttt, \yyVV, and \yytt decay channels with leptons in the final states; the total branching fraction is around 6.5\%.
The \bbll\ search targets final states arising from \HH decay channels where one of the Higgs bosons decays to a $b$-quark pair and the other to either a boson pair ($ZZ^{*}$, $WW^{*}$) or a $\tau$-lepton pair, which then decays to a pair of opposite-sign leptons ($\ell$ = $e$, $\mu$) and neutrinos, for a total branching fraction of 2.9\%.
Depending on the analysis, the final discriminating variable can be the \HH invariant mass, the diphoton invariant mass, or the multivariate classifiers used to separate signal from background.


\label{sec:detector}

\newcommand{\AtlasCoordFootnote}{
ATLAS uses a right-handed coordinate system with its origin at the nominal interaction point (IP)
in the center of the detector and the \(z\)-axis along the beam pipe.
The \(x\)-axis points from the IP to the center of the LHC ring,
and the \(y\)-axis points upwards.
Polar coordinates \((r,\phi)\) are used in the transverse plane,
\(\phi\) being the azimuthal angle around the \(z\)-axis.
The pseudorapidity is defined in terms of the polar angle \(\theta\) as \(\eta = -\ln \tan(\theta/2)\).
}

The analyses under consideration use the full sample of $\sqrt{s}$ = 13~\TeV proton-proton ($pp$) collision data recorded with the ATLAS detector during \RunTwo of the LHC. The integrated luminosity ranges from 126 to 140~\ifb\ depending on the trigger selection~\cite{TRIG-2016-01}.
The ATLAS experiment is a multipurpose particle detector with a forward-backward symmetric cylindrical geometry and nearly $4\pi$ coverage in solid angle~\cite{PERF-2007-01, ATLAS-TDR-19, PIX-2018-001}.
A software suite~\cite{SOFT-2022-02} is used in data simulation, in the reconstruction and analysis of real and simulated data, in detector operations, and in the trigger and data acquisition systems of the experiment.
The searches use a common set of event generators to describe ggF and VBF \HH production in the $pp$ collisions.
Reweighting methods are used to estimate the total and differential signal yields at a given value of $\kappa_{i}$ from samples simulated for different values of $\kl$ and $\kvv$~\cite{HDBS-2022-03} or to estimate the particle-level $m_{HH}$ distributions for alternative values of the Wilson coefficients using parameters from Ref.~\cite{alasfar2024effective}.


\label{sec:statmodel}

The results are derived from a likelihood function $L(\bm{\alpha},\bm{\theta})$, where $\bm{\alpha}$ denotes the vector of parameters of interest (POIs) in the statistical model and $\bm{\theta}$ is a set of nuisance parameters (NPs), including systematic uncertainty contributions and background parameters. This global likelihood function is the product of individual search likelihoods.
The profile-likelihood-ratio test statistic $-2\ln{\Lambda(\bm{\alpha},\bm{\theta})}=-2\ln{\frac{L\left(\bm{\alpha},\hat{\hat{\bm{\theta}}}(\bm{\alpha})\right)}{L\left(\hat{\bm{\alpha}},\hat{\bm{\theta}}\right)}}$ is used to determine the 68\% and 95\% CL intervals and local significance in the asymptotic approximation~\cite{Cowan:2010js}.
The $\mathrm{CL_{s}}$ method~\cite{Read:2002hq} is utilized to derive upper limits on the \HH production cross section. To evaluate the expected limits, Asimov datasets~\cite{Cowan:2010js} are generated, setting all NPs to their best-fit values in data and fixing the POIs to those posited in the hypothesis under test. The event samples from the combined searches are scrutinized for overlaps in both real and simulated data; they are found to be less than 1\% in the signal regions, and, thus, considered negligible.
 
Complete discussions of the systematic uncertainties considered in the individual searches are provided in Refs.~\cite{HDBS-2019-27,HDBS-2021-10,HDBS-2019-29,HDBS-2022-02,HDBS-2019-04,HDBS-2019-02}.
Correlations of these uncertainties between different searches are investigated. Uncertainties related to the data-taking conditions, such as those associated with the integrated luminosity and the mismodeling of the multiple $pp$ interactions per bunch crossing, are assumed to be correlated across the searches.
An exception is the integrated luminosity uncertainty in the resolved \bbbb analysis~\cite{HDBS-2019-29}, which employs a different calibration version.
Where applicable, uncertainties associated with physics objects common to two or more searches are considered correlated.
Correlations are also assumed for theoretical uncertainties affecting simulated signal and background processes, such as uncertainties in the QCD scale, proton parton distribution functions, and Higgs boson decay branching fractions. Systematic uncertainties that significantly influence the individual searches but are strongly constrained or pulled in the data fitting are treated as uncorrelated to prevent undue influence on the other searches. However, the impact of treating them as correlated or uncorrelated in the combination was checked and found to be negligible.


\label{sec:result}
 
\label{sec:result_mu}

The signal strength \muHH is defined as the ratio of the measured inclusive ggF and VBF \HH production cross section to the SM prediction $\SMsigmaHH=32.8_{-7.2}^{\,+2.1}$~fb.
This  \muHH measure assumes that the relative ggF and VBF production cross sections, Higgs boson decay branching fractions, and relative kinematic distributions correspond to the SM predictions.
The fit to data indicates a value of $\muHH=0.5^{\,+1.2}_{-1.0}=0.5_{-0.8}^{\,+0.9}(\text{stat})_{-0.6}^{\,+0.7}(\text{syst})$, where \enquote{stat}\ and \enquote{syst}\ denote the statistical and systematic uncertainties, respectively. The result is compatible with the SM prediction, with a $p$ value of 0.64.
Assuming $\sigmaHH = \SMsigmaHH$, the expected value is $\muHH=1.0_{-1.0}^{\,+1.2}=1.0_{-0.9}^{\,+1.0}(\text{stat})_{-0.5}^{\,+0.7}(\text{syst})$.
The primary systematic uncertainty arises from an estimated uncertainty of 100\% in modeling the radiation of additional heavy-flavor jets in the ggF single-Higgs-boson background production process~\cite{TOPQ-2017-12, STDM-2012-11, TOPQ-2012-16, HIGG-2018-29, HIGG-2019-13}, affecting \muHH by 25\%. The observed (expected) significance of \muHH is 0.4 (1.0) standard deviations, with respect to the hypothesis of no \HH production. No significant \HH signal is observed above the expected background, and a 95\% CL upper limit of 2.9 is placed on \muHH.
If \HH production is absent, the expected 95\% CL upper limit is 2.4, and in the SM case ($\muHH=1$) the expected upper limit is 3.4.
The expected upper limit is 17\% lower than in the previous combination~\cite{HDBS-2022-03}: 13\% from improvements in the \bbtt, \bbyy and \bbbb analyses and an additional 4\% from the inclusion of the \ml and \bbll channels.
This combination provides the best expected sensitivity to the \HH production cross section to date. Figure~\ref{fig:muHHLimit} displays the limits from the individual searches and their combination~\cite{footnote1} highlighting the \bbtt channel as the one expected to constrain \muHH the most. The $p$ value for compatibility between the \muHH value measured in the combination and those measured in the individual searches is 0.16.
The observed and expected 95\% CL upper limits on \sigmaHH from the combination are 86 and 71~fb, respectively, derived in this case excluding theoretical uncertainties in the \HH production cross section.
 
\begin{figure}[tb]
\centering
\includegraphics[width=0.5\textwidth]{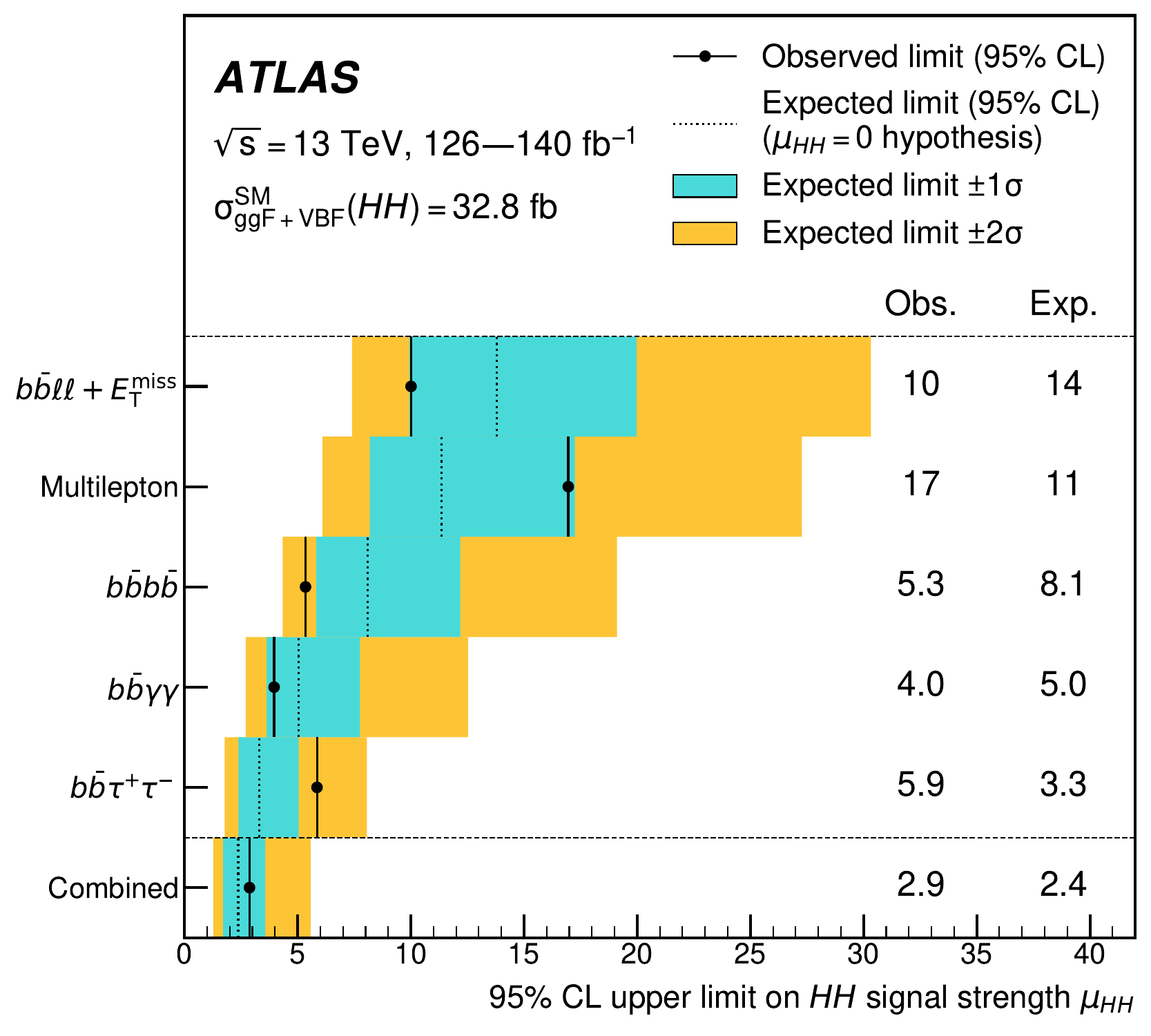}
\caption{Observed and expected 95\% CL upper limits on the signal strength for inclusive ggF \HH and VBF \HH production from the \bbtt, \bbyy, \bbbb, \ml and \bbll\ decay channels and their statistical combination. The predicted SM cross section assumes $m_H = 125$~\GeV. The expected limit, along with its associated $\pm1\sigma$ and $\pm2\sigma$ bands, is calculated for the assumption of no \HH production and with all NPs profiled to the observed data.}
\label{fig:muHHLimit}
\end{figure}


\label{sec:result_kappa}

The self-coupling modifier \kl is explored in the ggF and VBF \HH production processes.
The impact of \kl on the single-Higgs-boson background productions and the Higgs decay widths is neglected.
Assuming that other Higgs boson couplings conform to the SM predictions, a fit to data yields $\kl=3.8_{-3.6}^{\,+2.1}$, which is compatible with the SM prediction, with a $p$ value of 0.53.
The expected value of \kl is $1.0_{-1.5}^{\,+4.7}$ when assuming SM \HH production.
The observed (expected) 95\% CL interval is $-1.2<\kl<7.2$ ($-1.6<\kl<7.2$), representing the best expected sensitivity to the Higgs boson self-coupling to date. The values of the test statistic as a function of \kl are shown in Fig.~\ref{fig:kappaScan_a} for both the individual searches and their combination, highlighting the \bbyy channel as the most sensitive.
Similarly, \kvv is explored in the VBF \HH production process.
Assuming the SM predictions for other Higgs boson couplings, the observed (expected) value is $\kvv=1.02_{-0.23}^{\,+0.22}$ ($\kvv=1.00_{-0.36}^{\,+0.40}$).
The observed (expected) 95\% CL interval is $0.6<\kvv<1.5$ ($0.4<\kvv<1.6$).
The values of the test statistic as a function of \kvv are shown in Fig.~\ref{fig:kappaScan_b}, highlighting the \bbbb analysis as the most sensitive, mainly due to the boosted channel~\cite{HDBS-2022-02}. A deficit of data events in this channel results in stronger constraints on \kvv than expected.
To reduce model dependence, two-dimensional contours of $-2 \ln{\Lambda}$ in the \kvv--\kl plane are presented in Fig.~\ref{fig:kappaScan_c}. The $p$ value for compatibility of the combined measurement and the SM prediction is 0.78.
 
\begin{figure}[tb]
\centering
\subfloat[]{\includegraphics[width=0.5\textwidth]{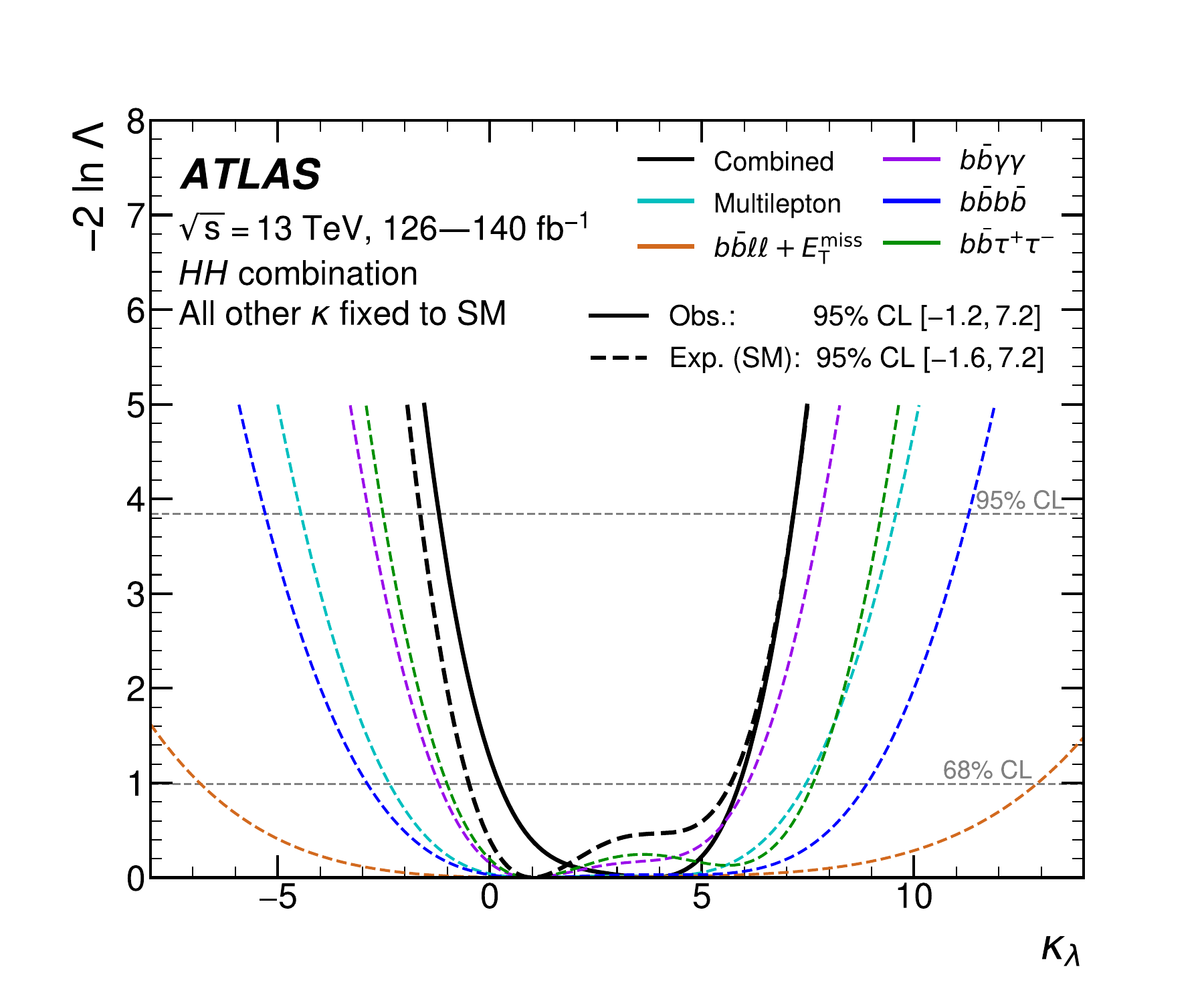}\label{fig:kappaScan_a}}
\subfloat[]{\includegraphics[width=0.5\textwidth]{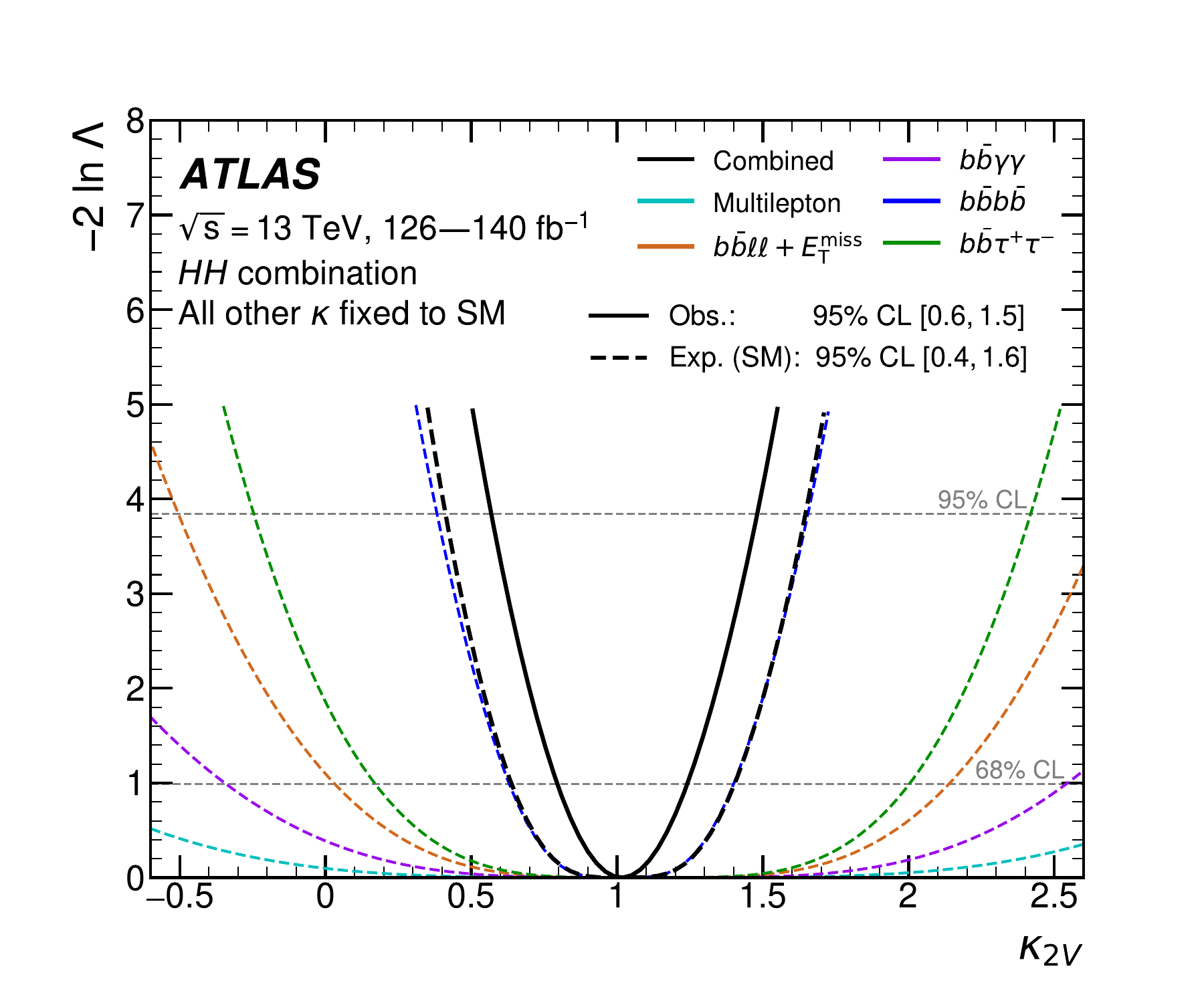}\label{fig:kappaScan_b}}\\
\subfloat[]{\includegraphics[width=0.5\textwidth]{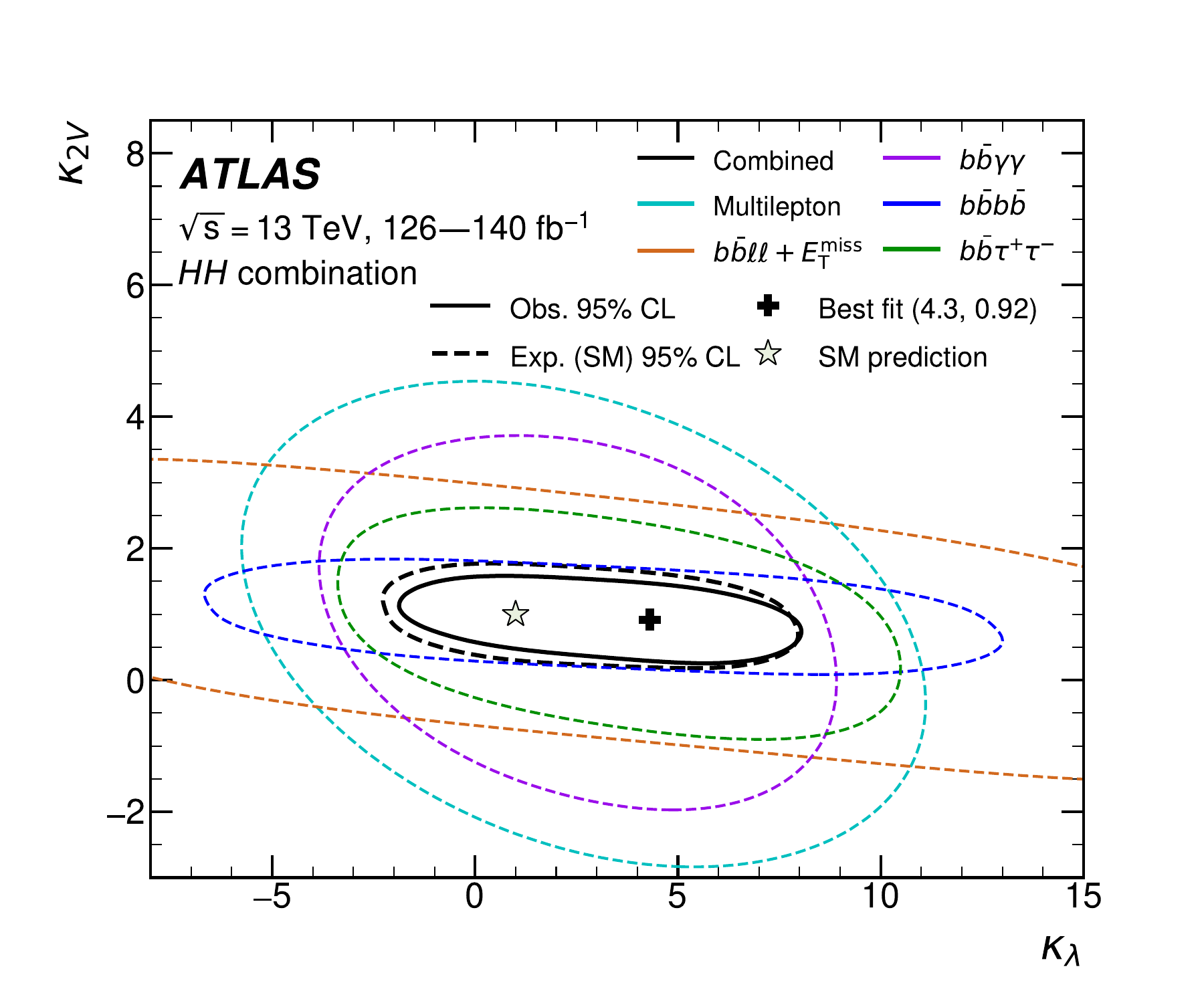}\label{fig:kappaScan_c}}
\caption{Expected values (dashed lines) of the test statistic ($-2\ln{\Lambda}$) as functions of (a) \kl and (b) \kvv. These results are shown for the decay channels \bbyy (purple), \bbtt (green), \ml (cyan), \bbbb (blue), and \bbll (brown), as well as their combination (black). The observed values from the combined data are depicted by solid black lines. These results are computed with the assumption that all other Higgs boson couplings follow the SM predictions.
(c) The expected 95\% CL contours in the \kvv-\kl plane, corresponding to the individual decay channels and their combination, are illustrated using dashed lines.
The observed contour from the combined results is depicted by a solid black line. The SM prediction is marked by a star, and the combined best-fit value is indicated by a cross.
}\label{fig:kappaScan}
\end{figure}


\label{sec:result_heft}

For the HEFT interpretation the three most sensitive \HH decay channels, \bbtt, \bbyy, and \bbbb, are combined.
The VBF \HH process is ignored, since it is sensitive only to \chhh and the predictions for this Wilson coefficient are not available for this process.
One-dimensional constraints are evaluated separately for the coefficients \cgghh and \ctthh, with all other coefficients fixed to the SM predictions.
At 95\% CL, the observed interval on \cgghh is $-0.38 < \cgghh < 0.49$; if the SM value $\cgghh=0$ is assumed, the expected 95\% CL interval is $-0.36<\cgghh<0.36$.
Similarly, the observed (expected) 95\% CL interval on \ctthh is $-0.19 < \ctthh < 0.70$ ($-0.27 < \ctthh < 0.66$). These represent the most stringent constraints to date on \cgghh and \ctthh. The results are compatible with the SM predictions, with $p$ values of 0.087 and 0.16, respectively.
Figure~\ref{fig:HEFT2D} displays the two-dimensional test-statistic contours in the coefficient spaces of (\cgghh, \chhh), (\ctthh, \chhh), and (\cgghh, \ctthh), with each plot fixing \ctthh, \cgghh, or \chhh, respectively, to its SM value. Two minima are expected because of the quadratic dependence of the cross section on the coefficients. The $p$ values for compatibility of the \cgghh-\chhh, \ctthh-\chhh, and \cgghh-\ctthh measurements with the SM predictions are 0.044, 0.21 and 0.031, respectively. The relatively low $p$ values are primarily due to the \bbbb analysis~\cite{HDBS-2019-29}, where the data-driven background modeling cannot perfectly describe the background distribution in data, making non-SM signals more favorable in the fit. Because of insufficient sensitivity, the combination does not allow simultaneous constraints to be placed on \chhh, \cgghh, and \ctthh in a more model-independent manner~\cite{footnote2}.
 
\begin{figure}[tb]
\centering
\subfloat[]{\includegraphics[width=0.5\textwidth]{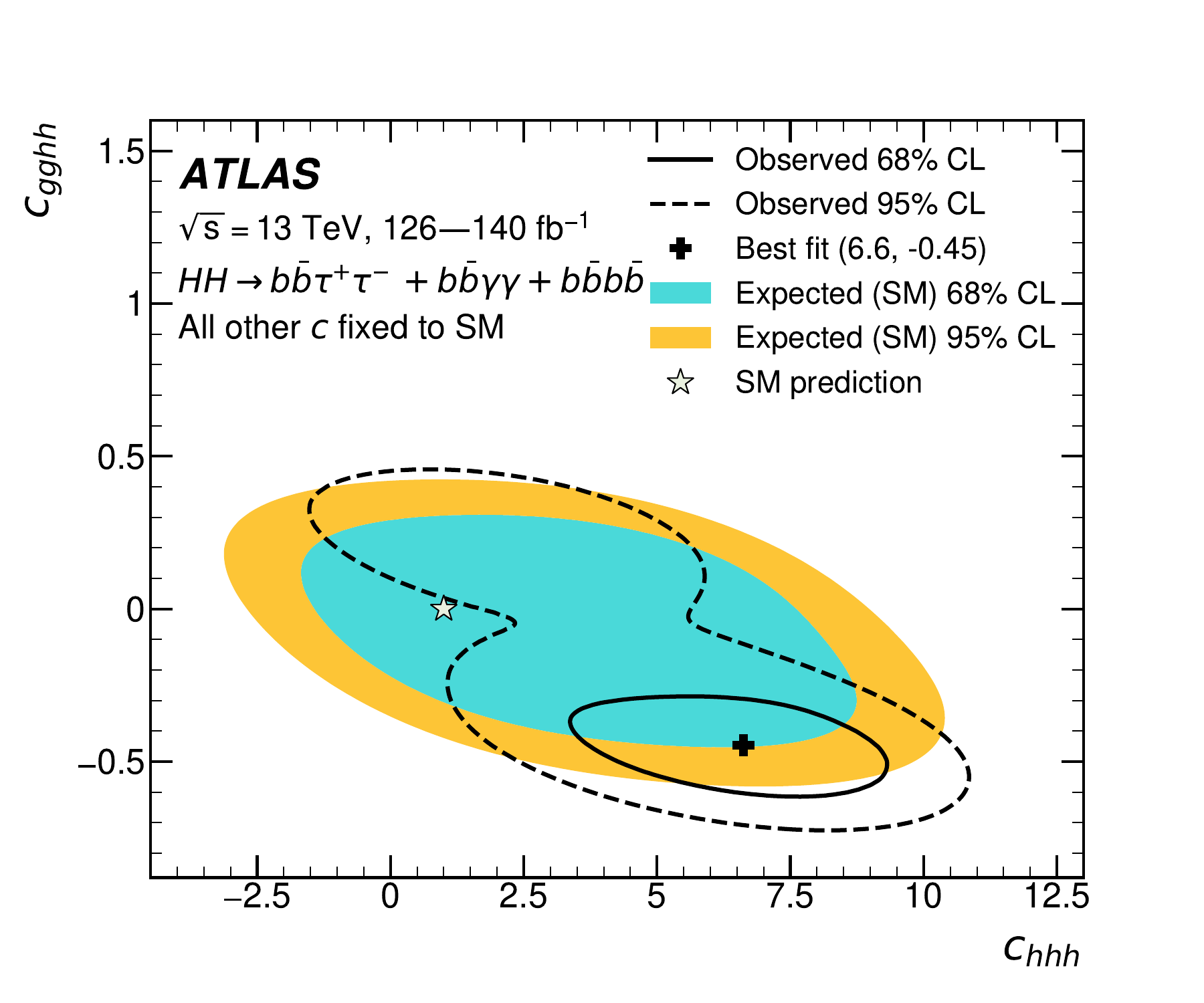}}
\subfloat[]{\includegraphics[width=0.5\textwidth]{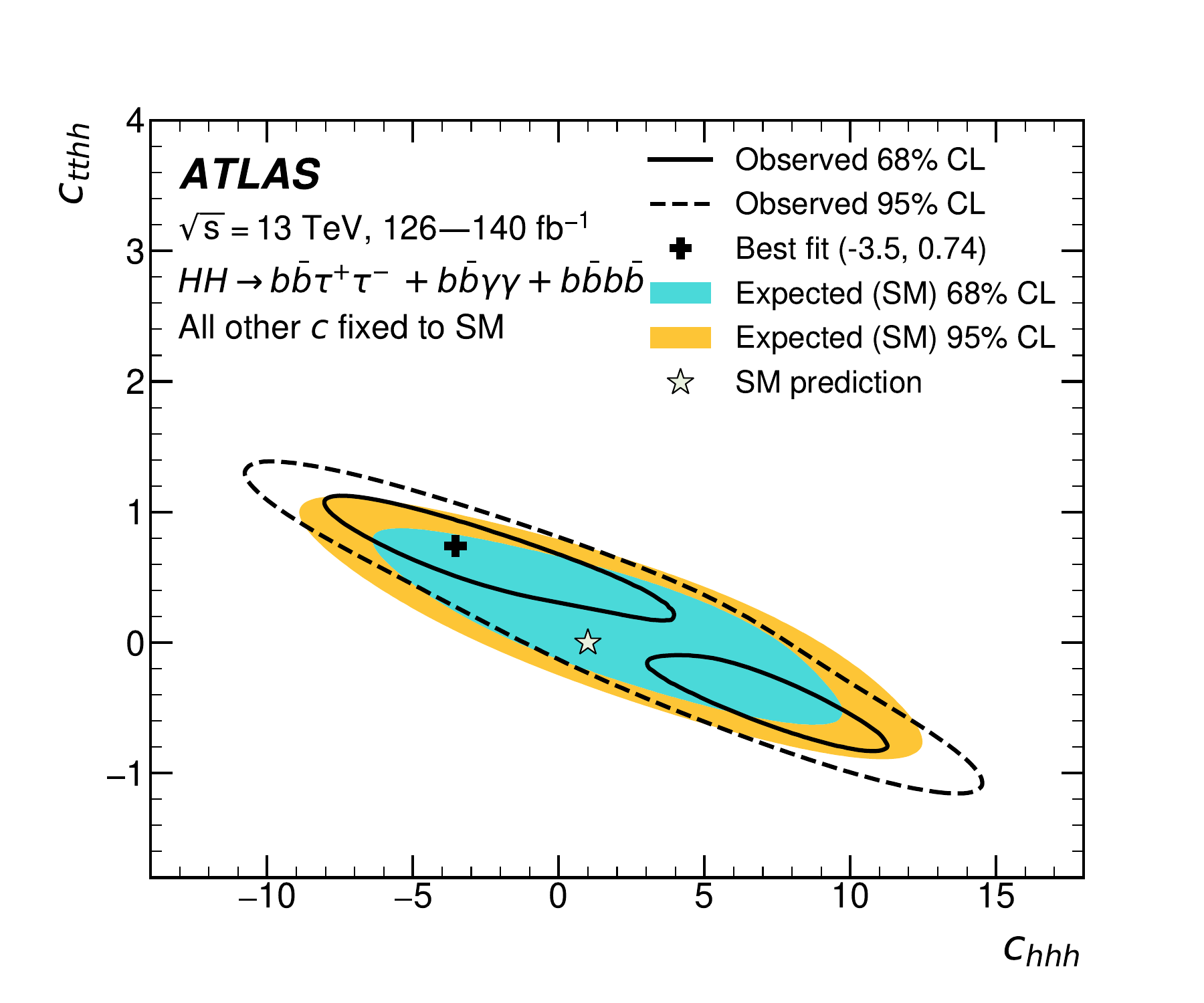}}\\
\subfloat[]{\includegraphics[width=0.5\textwidth]{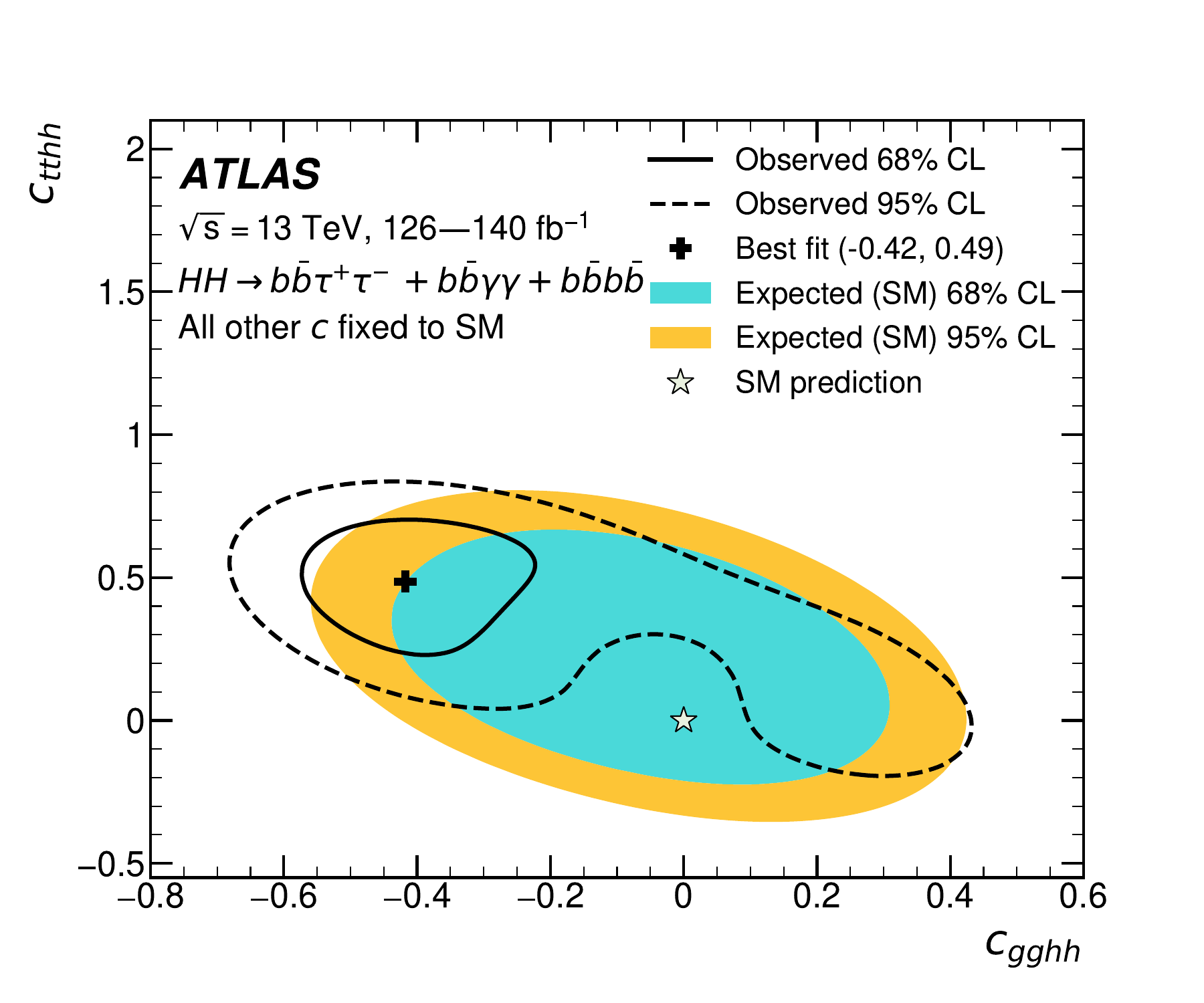}}
\caption{Two-dimensional test-statistic contours at 68\% CL (solid line) and 95\% CL (dashed line) in the (a) \cgghh-\chhh, (b) \ctthh-\chhh, and (c) \ctthh-\cgghh HEFT parameter spaces, with \ctthh, \cgghh, and \chhh fixed to their SM values, respectively. The corresponding SM expected contours are shown by the inner and outer shaded regions. The SM prediction is indicated by the star, while the best-fit value is shown by the cross.
}
\label{fig:HEFT2D}
\end{figure}


\label{sec:conclusion}

In summary, this Letter presents a combination of the results of searches for \HH production in the \bbtt, \bbyy, \bbbb, \ml, and \bbll decay channels, utilizing the complete LHC \RunTwo dataset of 13~\TeV proton-proton collisions recorded with the ATLAS detector.
This new combination provides the best expected sensitivities to the \HH production cross section and the Higgs boson self-coupling, superseding the results on the di-Higgs measurements of Ref.~\cite{HDBS-2022-03}. The results agree well with the SM predictions.
When using Higgs effective field theory to interpret the measurements, unprecedented constraints are placed on the effective $ggHH$ and $t\bar{t}HH$ interactions.


\FloatBarrier
\section*{\textit{Acknowledgments}---}


We thank CERN for the very successful operation of the LHC and its injectors, as well as the support staff at
CERN and at our institutions worldwide without whom ATLAS could not be operated efficiently.
 
The crucial computing support from all WLCG partners is acknowledged gratefully, in particular from CERN, the ATLAS Tier-1 facilities at TRIUMF/SFU (Canada), NDGF (Denmark, Norway, Sweden), CC-IN2P3 (France), KIT/GridKA (Germany), INFN-CNAF (Italy), NL-T1 (Netherlands), PIC (Spain), RAL (UK) and BNL (USA), the Tier-2 facilities worldwide and large non-WLCG resource providers. Major contributors of computing resources are listed in Ref.~\cite{ATL-SOFT-PUB-2023-001}.
 
We gratefully acknowledge the support of ANPCyT, Argentina; YerPhI, Armenia; ARC, Australia; BMWFW and FWF, Austria; ANAS, Azerbaijan; CNPq and FAPESP, Brazil; NSERC, NRC and CFI, Canada; CERN; ANID, Chile; CAS, MOST and NSFC, China; Minciencias, Colombia; MEYS CR, Czech Republic; DNRF and DNSRC, Denmark; IN2P3-CNRS and CEA-DRF/IRFU, France; SRNSFG, Georgia; BMBF, HGF and MPG, Germany; GSRI, Greece; RGC and Hong Kong SAR, China; ISF and Benoziyo Center, Israel; INFN, Italy; MEXT and JSPS, Japan; CNRST, Morocco; NWO, Netherlands; RCN, Norway; MNiSW, Poland; FCT, Portugal; MNE/IFA, Romania; MESTD, Serbia; MSSR, Slovakia; ARIS and MVZI, Slovenia; DSI/NRF, South Africa; MICINN, Spain; SRC and Wallenberg Foundation, Sweden; SERI, SNSF and Cantons of Bern and Geneva, Switzerland; NSTC, Taipei; TENMAK, T\"urkiye; STFC, United Kingdom; DOE and NSF, United States of America.
 
Individual groups and members have received support from BCKDF, CANARIE, CRC and DRAC, Canada; CERN-CZ, FORTE and PRIMUS, Czech Republic; COST, ERC, ERDF, Horizon 2020, ICSC-NextGenerationEU and Marie Sk{\l}odowska-Curie Actions, European Union; Investissements d'Avenir Labex, Investissements d'Avenir Idex and ANR, France; DFG and AvH Foundation, Germany; Herakleitos, Thales and Aristeia programmes co-financed by EU-ESF and the Greek NSRF, Greece; BSF-NSF and MINERVA, Israel; NCN and NAWA, Poland; La Caixa Banking Foundation, CERCA Programme Generalitat de Catalunya and PROMETEO and GenT Programmes Generalitat Valenciana, Spain; G\"{o}ran Gustafssons Stiftelse, Sweden; The Royal Society and Leverhulme Trust, United Kingdom.
 
In addition, individual members wish to acknowledge support from Armenia: Yerevan Physics Institute (FAPERJ); CERN: European Organization for Nuclear Research (CERN PJAS); Chile: Agencia Nacional de Investigaci\'on y Desarrollo (FONDECYT 1230812, FONDECYT 1230987, FONDECYT 1240864); China: Chinese Ministry of Science and Technology (MOST-2023YFA1605700), National Natural Science Foundation of China (NSFC - 12175119, NSFC 12275265, NSFC-12075060); Czech Republic: Czech Science Foundation (GACR - 24-11373S), Ministry of Education Youth and Sports (FORTE CZ.02.01.01/00/22\_008/0004632), PRIMUS Research Programme (PRIMUS/21/SCI/017); EU: H2020 European Research Council (ERC - 101002463); European Union: European Research Council (ERC - 948254, ERC 101089007), Horizon 2020 Framework Programme (MUCCA - CHIST-ERA-19-XAI-00), European Union, Future Artificial Intelligence Research (FAIR-NextGenerationEU PE00000013), Italian Center for High Performance Computing, Big Data and Quantum Computing (ICSC, NextGenerationEU); France: Agence Nationale de la Recherche (ANR-20-CE31-0013, ANR-21-CE31-0013, ANR-21-CE31-0022, ANR-22-EDIR-0002), Investissements d'Avenir Labex (ANR-11-LABX-0012); Germany: Baden-Württemberg Stiftung (BW Stiftung-Postdoc Eliteprogramme), Deutsche Forschungsgemeinschaft (DFG - 469666862, DFG - CR 312/5-2); Italy: Istituto Nazionale di Fisica Nucleare (ICSC, NextGenerationEU), Ministero dell'Università e della Ricerca (PRIN - 20223N7F8K - PNRR M4.C2.1.1); Japan: Japan Society for the Promotion of Science (JSPS KAKENHI JP22H01227, JSPS KAKENHI JP22H04944, JSPS KAKENHI JP22KK0227, JSPS KAKENHI JP23KK0245); Netherlands: Netherlands Organisation for Scientific Research (NWO Veni 2020 - VI.Veni.202.179); Norway: Research Council of Norway (RCN-314472); Poland: Polish National Agency for Academic Exchange (PPN/PPO/2020/1/00002/U/00001), Polish National Science Centre (NCN 2021/42/E/ST2/00350, NCN OPUS nr 2022/47/B/ST2/03059, NCN UMO-2019/34/E/ST2/00393, UMO-2020/37/B/ST2/01043, UMO-2021/40/C/ST2/00187, UMO-2022/47/O/ST2/00148, UMO-2023/49/B/ST2/04085); Slovenia: Slovenian Research Agency (ARIS grant J1-3010); Spain: Generalitat Valenciana (Artemisa, FEDER, IDIFEDER/2018/048), Ministry of Science and Innovation (MCIN \& NextGenEU PCI2022-135018-2, MICIN \& FEDER PID2021-125273NB, RYC2019-028510-I, RYC2020-030254-I, RYC2021-031273-I, RYC2022-038164-I), PROMETEO and GenT Programmes Generalitat Valenciana (CIDEGENT/2019/027); Sweden: Swedish Research Council (Swedish Research Council 2023-04654, VR 2018-00482, VR 2022-03845, VR 2022-04683, VR 2023-03403, VR grant 2021-03651), Knut and Alice Wallenberg Foundation (KAW 2018.0157, KAW 2018.0458, KAW 2019.0447, KAW 2022.0358); Switzerland: Swiss National Science Foundation (SNSF - PCEFP2\_194658); United Kingdom: Leverhulme Trust (Leverhulme Trust RPG-2020-004), Royal Society (NIF-R1-231091); United States of America: U.S. Department of Energy (ECA DE-AC02-76SF00515), Neubauer Family Foundation.


\clearpage

\printbibliography
 
\clearpage
 
\begin{flushleft}
\hypersetup{urlcolor=black}
{\Large The ATLAS Collaboration}

\bigskip

\AtlasOrcid[0000-0002-6665-4934]{G.~Aad}$^\textrm{\scriptsize 104}$,
\AtlasOrcid[0000-0001-7616-1554]{E.~Aakvaag}$^\textrm{\scriptsize 17}$,
\AtlasOrcid[0000-0002-5888-2734]{B.~Abbott}$^\textrm{\scriptsize 123}$,
\AtlasOrcid[0000-0002-0287-5869]{S.~Abdelhameed}$^\textrm{\scriptsize 119a}$,
\AtlasOrcid[0000-0002-1002-1652]{K.~Abeling}$^\textrm{\scriptsize 56}$,
\AtlasOrcid[0000-0001-5763-2760]{N.J.~Abicht}$^\textrm{\scriptsize 50}$,
\AtlasOrcid[0000-0002-8496-9294]{S.H.~Abidi}$^\textrm{\scriptsize 30}$,
\AtlasOrcid[0009-0003-6578-220X]{M.~Aboelela}$^\textrm{\scriptsize 45}$,
\AtlasOrcid[0000-0002-9987-2292]{A.~Aboulhorma}$^\textrm{\scriptsize 36e}$,
\AtlasOrcid[0000-0001-5329-6640]{H.~Abramowicz}$^\textrm{\scriptsize 154}$
% ...
\bigskip
\\

% ...
$^{17}$Department for Physics and Technology, University of Bergen, Bergen; Norway.\\
% ...
$^{30}$Physics Department, Brookhaven National Laboratory, Upton NY; United States of America.\\
% ...
$^{36}$$^{(a)}$Facult\'e des Sciences Ain Chock, Universit\'e Hassan II de Casablanca;$^{(b)}$Facult\'{e} des Sciences, Universit\'{e} Ibn-Tofail, K\'{e}nitra;$^{(c)}$Facult\'e des Sciences Semlalia, Universit\'e Cadi Ayyad, LPHEA-Marrakech;$^{(d)}$LPMR, Facult\'e des Sciences, Universit\'e Mohamed Premier, Oujda;$^{(e)}$Facult\'e des sciences, Universit\'e Mohammed V, Rabat;$^{(f)}$Institute of Applied Physics, Mohammed VI Polytechnic University, Ben Guerir; Morocco.\\
% ...
$^{45}$Physics Department, Southern Methodist University, Dallas TX; United States of America.\\
% ...
$^{50}$Fakult\"{a}t Physik , Technische Universit{\"a}t Dortmund, Dortmund; Germany.\\
% ...
$^{56}$II. Physikalisches Institut, Georg-August-Universit\"{a}t G\"ottingen, G\"ottingen; Germany.\\
% ...
$^{104}$CPPM, Aix-Marseille Universit\'e, CNRS/IN2P3, Marseille; France.\\
% ...
$^{119}$$^{(a)}$New York University Abu Dhabi, Abu Dhabi;$^{(b)}$United Arab Emirates University, Al Ain; United Arab Emirates.\\
% ...
$^{123}$Homer L. Dodge Department of Physics and Astronomy, University of Oklahoma, Norman OK; United States of America.\\
% ...
$^{154}$Raymond and Beverly Sackler School of Physics and Astronomy, Tel Aviv University, Tel Aviv; Israel.\\
% ...
$^{a}$ Also Affiliated with an institute covered by a cooperation agreement with CERN.\\
$^{b}$ Also at An-Najah National University, Nablus; Palestine.\\
$^{c}$ Also at Borough of Manhattan Community College, City University of New York, New York NY; United States of America.\\
$^{d}$ Also at Center for Interdisciplinary Research and Innovation (CIRI-AUTH), Thessaloniki; Greece.\\
$^{e}$ Also at CERN, Geneva; Switzerland.\\
$^{f}$ Also at CMD-AC UNEC Research Center, Azerbaijan State University of Economics (UNEC); Azerbaijan.\\
% ...

\end{flushleft}


\end{document}